\documentclass[12pt]{article}

\usepackage{graphicx}
\usepackage{amssymb}
\usepackage{amsfonts}
\usepackage{amsmath}
\usepackage{epsfig}
\usepackage{epsf,hyperref}

\begin{document}

\title{Modular quantum computing and quantum-like devices}
\author{\textit{R.Vilela Mendes}\thanks{%
rvilela.mendes@gmail.com; rvmendes@fc.ul.pt; http://label2.ist.utl.pt/vilela/%
} \\
CMAFCIO, Faculdade de Ci\^{e}ncias, Universidade de Lisboa}
\date{ }
\maketitle

\begin{abstract}
The two essential ideas in this paper are, on the one hand, that a
considerable amount of the power of quantum computation may be obtained by
adding to a classical computer a few specialized quantum modules and, on the
other hand, that such modules may be constructed out of classical systems
obeying quantum-like equations where a space coordinate is the evolution
parameter (thus playing the role of time in the quantum algorithms).
\end{abstract}


\section{Introduction}

\subsection{Computation models}

Classical, probabilistic and quantum computing are three computing
modalities which, adopting a Turing Machine-like scheme \cite{Deutsch} \cite%
{Bernstein}, may be briefly described in the following way:

Let $M$ be a states machine with one working tape with alphabet $\Gamma $
and an input tape with alphabet $\Sigma $. At each time the machine
configuration $c$ is the content of the working tape, the position of two
pointers (in the input and working tapes) and the current state. Let $%
\mathcal{C}\left( x\right) $, of cardinality $N$, be the set of all possible
configurations when the input is $x$.

At each time step the machine, in a state $q\in Q$, reads a symbol $\sigma
\in \Sigma $ in the input tape and the current symbol $\gamma \in \Gamma $
in the working tape, changes to a state $q^{\prime }\in Q$, prints a symbol $%
\gamma ^{\prime }\in \Gamma $ in the working tape and the pointers move
right ($R$) or left ($L$) in the respective tapes. The probability of these
operations is controlled by a mapping $T$ from $\mathcal{C}\left( x\right) $
into a space $S$%
\begin{equation*}
T:Q\times \Sigma \times \Gamma \times Q\times \Gamma \times
\{L,R\}^{2}\rightarrow S
\end{equation*}%
This mapping is called the transition function from which the transition
probability between $c_{i}$ and the next $c_{i+1}$ configuration $p\left(
c_{i},c_{i+1}\right) =F\left( T\left( c_{i},c_{i+1}\right) \right) $ may be
obtained. In all cases it is assumed that the internal state of the machine
is not observed except at the final time of the calculation. The mapping $T$
defines a matrix in the space of configurations $\mathcal{C}$.%
\begin{equation*}
T\left( q_{j},\gamma _{j},p_{j}^{(1)}p_{j}^{(2)}|q_{i},\gamma
_{i},p_{i}^{(1)}p_{i}^{(2)}\right) \circeq T\left( c_{j},c_{i}\right)
\end{equation*}

The three computation models correspond to different choices of $T$ and of $%
p\left( c_{i},c_{i+1}\right) =F\left( T\left( c_{i},c_{i+1}\right) \right) $

\textit{Classic deterministic computation:}%
\begin{eqnarray}
S &=&\left\{ s:s=0,1\right\}  \notag \\
p\left( c_{j},c_{i}\right) &=&T\left( c_{j},c_{i}\right) =s  \label{1.1}
\end{eqnarray}%
Only one element in each line of the transition matrix $T$ is different from
zero.

\textit{Classical probabilistic computation:}%
\begin{eqnarray}
S &=&\left\{ s:s\in \left[ 0,1\right] \right\}  \notag \\
p\left( c_{j},c_{i}\right) &=&T\left( c_{j},c_{i}\right) =s  \label{1.2}
\end{eqnarray}%
with the condition%
\begin{equation}
\sum_{j}T\left( c_{j},c_{i}\right) =1  \label{1.3}
\end{equation}%
$T$ is a stochastic matrix preserving the $\mathcal{L}^{1}$ norm in the
space of configurations.

\textit{Quantum computation:}%
\begin{eqnarray}
S &=&\left\{ s\in \mathbb{C}:\left\vert s\right\vert ^{2}=1\right\}  \notag
\\
p\left( c_{j},c_{i}\right) &=&\left\vert T\left( c_{j},c_{i}\right)
\right\vert ^{2}=\left\vert s\right\vert ^{2}  \label{1.4}
\end{eqnarray}%
with the condition%
\begin{equation}
\sum_{j}\left\vert T\left( c_{j},c_{i}\right) \right\vert ^{2}=1  \label{1.5}
\end{equation}%
that is, $T$ is a unitary matrix preserving the $\mathcal{L}^{2}$ norm in
the space of configurations.

In all cases the transition probabilities between initial and final states
are positive and normalized. The difference between the three computational
models is the method used to find the transition probabilities.

Physical implementations of the computational models require physical
elements for \textbf{coding}, \textbf{interaction} between the elements to
perform the writing and change of states and finally an \textbf{evolution}
process to represent the transition function. \textit{Coding, interaction
and evolution}. And, in each case, the evolution should be such as to
satisfy the constraints (\ref{1.1}) or (\ref{1.3}) or (\ref{1.5}).

Some quantum systems, when sufficiently isolated from the environment,
because their coherent time-evolution is unitary, provide physical models of
quantum computation. However, quantum computation is not quantum mechanics.
Any other system, that provides coding, interaction and a change of states
compatible with (\ref{1.4}) (\ref{1.5}), may also provide a model of quantum
computation. In particular the state evolution of these systems should be
unitary. Such systems have been called quantum-like.

In Ref.\cite{Manko1} it has been proposed that classical paraxial light
propagation, being ruled by a Schr\"{o}dinger-like equation may also provide
a model of quantum computation. There is, of course, no contradiction with
the physical rules of quantum mechanics because in the classical paraxial
system the propagation is along a space coordinate which plays the same role
as time in the quantum mechanical Schr\"{o}dinger equation. As a consequence
the transfer function may be implemented by the unitary propagation of
information along a space coordinate. Considering the coding and interaction
requirements, a good candidate for this implementation seems to be fiber or
wave-guide optics.

The idea of using quantum-like systems for quantum computation and
simulation of quantum effects has been later explored (see for example \cite%
{Moya1} - \cite{Moya4}) by several authors.

\subsection{Modular computation}

Although it has not yet been rigorously proven that $BPP\subsetneq BQP$,
that is, that quantum circuits cannot be efficiently simulated in a
bounded-error probabilistic machine, the quantum oracle algorithms, that
have been developed, provide circumstantial evidence that quantum computing
is indeed more efficient than classical computing.

The power of quantum computing hinges both on the capacity to deal with
superpositions of many different states (quantum parallelism) and on the
enhancement of particular computational paths (quantum interference). The
following three resources are responsible for the efficiency of the known
quantum algorithms:

(i) Preparation of a linear superposition of all possible basis states $%
\sum_{x}\left| x\right\rangle $;

(ii) Call to a reversible oracle operation 
\begin{equation*}
\sum_{x}\left\vert x\right\rangle \left\vert \psi \right\rangle \rightarrow
\sum_{x}\left\vert x\right\rangle \left\vert f\left( x\right) \oplus \psi
\right\rangle =\sum_{x}\left\vert x\right\rangle U_{f\left( x\right)
}\left\vert \psi \right\rangle
\end{equation*}%
the target qubit(s) $\left\vert \psi \right\rangle $ being usually chosen to
be eigenstates of the controlled unitary operations $U_{f\left( x\right) }$
with eigenvalues $e^{i\alpha \left( x\right) }$;

(iii) Use of the $e^{i\alpha \left( x\right) }$ phases (kicked back to $%
\left| x\right\rangle $) to enhance, by interference, particular
computational paths.

The oracle is the quantum subroutine that contains the information specific
to each particular problem. The way the oracle is chosen to act (in
particular the choice of the target qubit as an eigenstate of $U_{f\left(
x\right) }$) implies that the natural interference device is the quantum
Fourier transform (QFT). On the other hand, the QFT,\ operating on the state 
$\left\vert 00\cdots 0\right\rangle $, also generates a superposition of all
the basis states. This suggests that most of the power of quantum computing
may be obtained by adding to a classical computer a few basic modules,
namely:

(i) A quantum Fourier transform module

(ii) Programmable oracle modules.

In theoretical discussions the oracle is considered to be a subroutine call,
invocation of which only costs unit time. However, one should not forget
that it is an operation acting in all basis states and therefore, to benefit
from quantum parallelism the practical requirements for its implementation
are not very different from those of the quantum Fourier transform.

Quantum computing requires the coding, manipulation and detection of
entangled qubits. Nuclear spins, atom states, flux units, Cooper pairs or
single photon polarizations have been proposed and used to encode qubits and
exhibit quantum logic operations. Qubits encoded in such fundamental matter
units might indeed be the ultimate building blocks of future quantum
computers. For practical computing applications, a scalable tensor product
structure is required to avoid an exponential demand for physical resources.
However, this seems difficult to achieve with the prototype quantum gates
that have been developed. Therefore a search for alternative implementations
seems appropriate.

Section 2, improving and extending a previous proposal \cite{Manko1},
discusses an implementation of quantum computing operations in classical
systems that propagate according to a Schr\"{o}dinger equation with a space
coordinate playing the role of time. Here one tries to make a concrete
proposal for the implementation of the theory using fiber or wave-guide
optics, the qubits being robustly coded in particular modes or on their
polarizations, with the result of the (unitary) operations being read off at
particular locations of the optical systems. Fiber or planar wave-guide
optics implementations benefit from the large amount of technological
sophistication already developed for communications. Therefore, the emphasis
is on the construction of quantum gates using devices and techniques
currently available in this field. As the sophisticated optical elements
developed so far have been done mostly for telecommunication purposes one
also clarifies the implementation progress needed to make them appropriate
for the quantum computation purposes. As well as the issues of coding and
gate implementation, also polarization effects, signal coupling and the
notions of mixing, entanglement and coherence are discussed in this setting.

Finally, Section 3 discusses how these quantum-like elements might scale-up
to construct a quantum Fourier transform module as well as programmable
oracles.

\section{Quantum-like computation with fiber or wave-guide optics}

\subsection{Unitary evolution}

In optical fibers or planar wave-guides, mode propagation may be well
approximated by a Schr\"{o}dinger equation with the longitudinal $z$%
-coordinate playing the role of time. Ref.\cite{Manko1} follows a reasoning
similar to the Leontovich-Fock \cite{Fock} description of paraxial beams in
the parabolic approximation. Here one generalizes the derivation in \cite%
{Manko1} by explicitly including polarization effects.

From the Maxwell equations, with $\rho =J=M=0$, one obtains the Helmholz
equation for a space-varying dielectric constant 
\begin{equation}
\bigtriangledown \left( \frac{1}{\varepsilon }E\cdot \bigtriangledown
\varepsilon \right) +\bigtriangleup E=\varepsilon \mu _{0}\frac{\partial
^{2}E}{\partial t^{2}}  \label{2.1}
\end{equation}%
Consider now a fixed frequency transversal mode $E\left( x,y,z,t\right)
=E\left( x,y,z\right) \exp \left( i\omega t\right) $ and an index of
refraction profile 
\begin{equation}
\varepsilon \left( x,y,z\right) =n^{2}\left( x,y,z\right) =n_{0}^{2}\left(
z\right) -V\left( x,y\right)  \label{2.2}
\end{equation}%
where $n_{0}^{2}\left( z\right) $ is the index of refraction at the fiber
axis and $V\left( x,y\right) <<n_{0}^{2}\left( z\right) $. With this last
condition and neglecting terms in $\left( \bigtriangledown V\right) ^{2}$
and $V\bigtriangledown V$, one obtains, for a transversal electric mode 
\begin{equation}
\left( \frac{\partial ^{2}}{\partial z^{2}}+\bigtriangleup _{2}\right)
E+n^{2}k_{0}^{2}E-\frac{1}{n_{0}^{2}}\bigtriangledown _{2}\left( E\cdot
\bigtriangledown _{2}V\right) \simeq 0  \label{2.3}
\end{equation}%
where $k_{0}=\frac{\omega }{c}$ and $\lambda _{0}=\frac{2\pi }{k_{0}}$ is
the wavelength in vacuum.

Introduce the slowly varying (in $z$) complex vectorial function $\psi
\left( x,y,z\right) $%
\begin{equation}
E\left( x,y,z\right) =\psi \left( x,y,z\right) \exp \left(
ik_{0}\int^{z}n_{0}\left( \zeta \right) d\zeta \right)  \label{2.4}
\end{equation}%
For slow variation of the index of refraction along the fiber axis over
distances of the order of one wavelength%
\begin{equation*}
\frac{\lambda _{0}}{n_{0}^{2}\left( z\right) }\left\vert \frac{dn_{0}\left(
z\right) }{dz}\right\vert <<1
\end{equation*}%
one may neglect second-order derivatives of $\psi $ along $z$ and
derivatives of $n_{0}\left( z\right) $ and end up with 
\begin{equation}
i\lambda _{0}\frac{\partial }{\partial z}\left( 
\begin{array}{l}
\psi _{x} \\ 
\psi _{y}%
\end{array}%
\right) =\left\{ 
\begin{array}{c}
\frac{\lambda _{0}^{2}}{4n_{0}\left( z\right) }\left( 
\begin{array}{cc}
-\bigtriangleup _{2} & 0 \\ 
0 & -\bigtriangleup _{2}%
\end{array}%
\right) +\frac{\pi }{n_{0}\left( z\right) }\left( 
\begin{array}{cc}
V\left( x,y\right) & 0 \\ 
0 & V\left( x,y\right)%
\end{array}%
\right) \\ 
+\frac{\lambda ^{2}}{4\pi n_{0}^{3}\left( z\right) }\left( 
\begin{array}{cc}
\partial _{x}^{2}V+\partial _{x}V\partial _{x} & \partial
_{xy}^{2}V+\partial _{y}V\partial _{x} \\ 
\partial _{xy}^{2}V+\partial _{x}V\partial _{y} & \partial
_{y}^{2}V+\partial _{y}V\partial _{y}%
\end{array}%
\right)%
\end{array}%
\right\} \left( 
\begin{array}{l}
\psi _{x} \\ 
\psi _{y}%
\end{array}%
\right)  \label{2.5}
\end{equation}%
which is a quantumlike version of the Schr\"{o}dinger-Pauli equation. The
role of time in this equation is played by the spatial (longitudinal)
coordinate of the light beam, the role of Planck's constant is played by the
light wavelength and the role of potential energy by the index of refraction
of the medium. Thus, a beam of light, a purely classical object, obeys
equations formally identical to those of quantum mechanics.

The unitary $z$-evolution of the electromagnetic complex amplitude is
described by the evolution operator $\hat{U}(z)$ 
\begin{equation}
\hat{U}(z,z_{0})\psi (x,y,z_{0})=\psi (x,y,z),  \label{2.6}
\end{equation}%
associated to the Hamiltonian 
\begin{equation}
\hat{H}(z)=\left( \frac{\hat{p}_{x}^{2}}{2}+\frac{\hat{p}_{y}^{2}}{2}\right) 
\frac{1}{n_{0}(z)}+\Gamma (x,y,z).  \label{2.7}
\end{equation}%
with $\hat{p}_{x}=-i\lambda \,\frac{\partial }{\partial x}\,,\hat{p}%
_{y}=-i\lambda \,\frac{\partial }{\partial y}\,$ and a potential function $%
\Gamma (x,y,z)$ which, for general $V\left( x,y\right) $, has local and
nonlocal terms mixing the polarizations as follows from Eq.(\ref{2.5}).
Manipulation of the polarization will play an important role in this
quantumlike computation approach. As seen from the last term in Eq.(\ref{2.5}%
) it is obtained by engineering the index of refraction profile.

Other quantumlike systems are reviewed in~\cite{QL1} \cite{QL2} \cite%
{Margarita}. They include sound-wave propagation in acoustic waveguides,
charged-particle beams and light beams inside diode lasers. Full
implementation of quantum algorithms might also be obtained in these
systems. For each unitary operation a steady state is to be established and
the result of the computation is read at the appropriate space location. The
notion of preservation of \textit{time coherence} needed to define the
reliability and maximum number of operations in quantum computation is here
replaced by \textit{space coherence} of the steady state that is established
in the device.

Before discussing practical implementations of the quantumlike
representation, I add two speculative remarks:

(i) Abrams and Lloyd \cite{Lloyd1} have shown that were quantum mechanics
nonlinear, more computational power could still be achieved. There is no
evidence indicating that actual quantum mechanics is nonlinear. However, in
the quantumlike scheme it is quite simple to implement a nonlinear Schr\"{o}%
dinger equation evolving in the $z-$coordinate. Therefore, quantumlike
nonlinear circuits might provide an adequate framework to test Abrams and
Lloyd's ideas.

(ii) Brun \cite{Timelike} has pointed out that hard problems could in
principle be solved, even by a classical computer, if it had access to a
closed timelike curve. Except maybe in extreme cosmological conditions,
closed timelike curves are not readily available. However, if in a
computational scheme (both classical and quantum) time is replaced by space,
simulation of closed timelike curves is not unthinkable.

Before proceeding it should be pointed out that other implementation of some
features of quantum algorithms by linear optical methods have been proposed
by several authors (see for example \cite{Ralph} \cite{Brien} and references
therein). To obtain the entanglement needed for universal quantum
computation, the proposed optics implementations use either:

(i) Kerr nonlinearities, which are hard to achieve at the single-photon
level or

(ii) a probabilistic scheme based on the nonlinearity implicit in the
selection by single-photon detectors.

What is proposed here and in Ref.\cite{Manko1} is a more radical proposal in
the sense that, instead of setting up a time sequence of optical events as
the implementation of the quantum algorithm, one uses the fact that, in
optical fibers, mode propagation is well approximated by a Schr\"{o}dinger
equation with the $z$-coordinate along the fiber playing the role of time.

\subsection{Coding}

In Ref.\cite{Manko1}, several ways to code qubits on a fiber, using either
discrete or continuous variables, were already discussed. Here simpler
implementations are proposed which might be robustly obtained with the
materials available for optical communication applications.

Consider three types of qubit codings in two types of fibers:

(a) In single-mode (double-polarization) fibers a qubit would correspond to
the two polarizations directions of the $LP_{01}$ mode \cite{Gloge} (Fig.\ref%
{Fig1}).

(b) In single-mode fibers a qubit might also be associated to the amplitudes
of a particular polarization in two distinct fibers, one of the fibers
associated to $\left\vert 0\right\rangle $ and the other to $\left\vert
1\right\rangle $.

\begin{figure}[htb]
\centering
\includegraphics[width=0.6\textwidth]{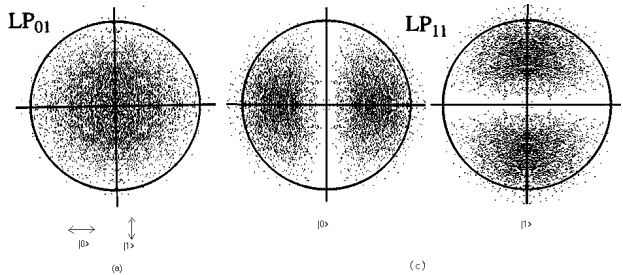}
\caption{Two distinct qubit coding choices}
\label{Fig1}
\end{figure}

(c) In fibers with normalized frequency allowing for $LP_{01}$ and $LP_{11}$
modes, a qubit may be associated to the two distinct $LP_{11}$ modes,
without distinguishing polarization states (Fig.\ref{Fig1}). Counting the
polarizations one has four degrees of freedom associated to the $LP_{11}$
mode, which allows for the coding of two qubits and the implementation of a
two-qubit gate in a single fiber (see below).

These codings are the simplest ones for optical fiber implementations.
Notice however that the reliability of fiber optics techniques allows for
reliable manipulation and separation of many other modes. For example for a
fiber with normalized frequency $5.5201<\frac{2\pi r_{co}}{\lambda }\sqrt{%
n_{co}^{2}-n_{cl}^{2}}<6.3802$, modes up $LP_{12}$ may be excited, allowing
for $40$ different quantumlike degrees of freedom in a single fiber. ($%
r_{co},n_{co},n_{cl}$ denote the core radius and the index of refraction of
core and cladding)

\subsection{One-qubit gates}

Universal quantum computation requires one-qubit gates performing arbitrary
unitary transformation and, at least, a two-qubit gate performing a unitary
transformation in the four-dimensional tensor space which, together with the
one-qubit transformations, generates the unitary group in four dimensions.
For the one-qubit gates, two schemes seem appropriate:

\textbf{(1) Polarization coding}

Isotropic single-mode fibers support two degenerate polarization modes which
propagate with the same constants $\beta _{i}=k_{0}n_{i}$ . However it is
relatively easy to make the fibers to behave as linearly birefringent or
circularly birefringent media \cite{Pollock} \cite{Kasap}. The birefringence
of the fiber is conventionally characterized by a Jones matrix $J$ \cite{J1} 
\cite{J2} \cite{Chen2} which defines the amount of transformation of the
phase induced by the difference $\Delta \beta _{L}=\beta _{x}-\beta _{y}$
(or $\Delta \beta _{C}=\beta _{R}-\beta _{L}$ for circular polarization).
Because a global phase associated to $k_{0}\int^{z}n_{0}\left( \zeta \right)
d\zeta $ is already taken into account, the Jones matrix is what defines the
phase rotation of $\psi \left( x,y,z\right) $ in Eq.(\ref{2.4}). For linear
birefringence the Jones matrix relating the output and input phase of a
fiber of length $L$, is 
\begin{equation}
J_{L}\left( \Delta \beta _{L}\right) =\left( 
\begin{array}{ll}
e^{iL\Delta \beta _{L}/2} & 0 \\ 
0 & e^{-iL\Delta \beta _{L}/2}%
\end{array}%
\right)  \label{2.10}
\end{equation}%
In this expression it is assumed that the fast axis, that is, the one with
the largest $\beta $, is in the $x-$direction. If the fast axis is at an
angle $\theta $ relative to the $x-$direction the Jones matrix would be 
\begin{equation}
J_{L}\left( \Delta \beta _{L},\theta \right) =\left( 
\begin{array}{ll}
\cos \frac{L\Delta \beta _{L}}{2}+i\cos 2\theta \sin \frac{L\Delta \beta _{L}%
}{2} & i\sin 2\theta \sin \frac{L\Delta \beta _{L}}{2} \\ 
i\sin 2\theta \sin \frac{L\Delta \beta _{L}}{2} & \cos \frac{L\Delta \beta
_{L}}{2}-i\cos 2\theta \sin \frac{L\Delta \beta _{L}}{2}%
\end{array}%
\right)  \label{2.11}
\end{equation}%
Any $U\left( 2\right) $ matrix may be decomposed into 
\begin{equation}
U\left( \alpha ,\theta ,\beta \right) =\left( 
\begin{array}{ll}
e^{i\alpha /2} & 0 \\ 
0 & e^{-i\alpha /2}%
\end{array}%
\right) \left( 
\begin{array}{ll}
\cos \frac{\theta }{2} & i\sin \frac{\theta }{2} \\ 
i\sin \frac{\theta }{2} & \cos \frac{\theta }{2}%
\end{array}%
\right) \left( 
\begin{array}{ll}
e^{i\beta /2} & 0 \\ 
0 & e^{-i\beta /2}%
\end{array}%
\right)  \label{2.12}
\end{equation}%
hence it follows from (\ref{2.10}) and (\ref{2.11}) that any $U\left(
2\right) $ transformation may be obtained on linearly birefringent fibers.

Linear birefringence is easily obtained by elliptical cores, lateral stress,
bending or application of an electrical field. For fixed one qubit gates the
most robust method is probably the use of cooling induced stress \cite%
{Eickhoff}. Circular birefringence is obtained by geometrical twisting (spun
fibers) or axial magnetic fields (Faraday rotation). Adjustment of the
intensity of these properties by the variation of applied electromagnetic
fields is a potentially useful feature for the construction of programmable
modules.

Engineering the birefringency properties is a very flexible way to obtain
one qubit gates using single mode double-polarization fibers. For example,
in the cases above one has assumed that the fast axis is fixed along the
fiber segment. If instead one has a continuously rotating fast axis, a more
complex Jones matrix is obtained%
\begin{equation*}
J_{L}\left( \Delta \beta _{L},\xi \right) =\left( 
\begin{array}{ll}
\cos \frac{\delta }{2}+i\frac{L\Delta \beta _{L}}{\delta }\sin \frac{\delta 
}{2} & \frac{L\xi }{\delta }\sin \frac{\delta }{2} \\ 
-\frac{L\xi }{\delta }\sin \frac{\delta }{2} & \cos \frac{\delta }{2}-i\frac{%
L\Delta \beta _{L}}{\delta }\sin \frac{\delta }{2}%
\end{array}%
\right)
\end{equation*}%
with $\delta =\sqrt{\left( L\Delta \beta _{L}\right) ^{2}+4\left( L\xi
\right) ^{2}}$ and $\xi =\frac{d\theta }{dz}$ the constant rate of rotation
of the fast axis along the $z-$coordinate.

Also, for a simple circularly birefringent fiber the Jones matrix is%
\begin{equation*}
J_{L}\left( \Delta \beta _{C}\right) =\left( 
\begin{array}{ll}
\cos \frac{L\Delta \beta _{C}}{2} & \sin \frac{L\Delta \beta _{C}}{2} \\ 
\sin \frac{L\Delta \beta _{C}}{2} & \cos \frac{L\Delta \beta _{C}}{2}%
\end{array}%
\right)
\end{equation*}%
and for a fiber that is both linearly and circularly birefringent (for
example a linearly birefringent spun fiber or a linearly birefringent one
with an axial magnetic field) the Jones matrix is%
\begin{equation*}
J\left( \Delta \beta _{C},\alpha \right) =\left( 
\begin{array}{ll}
\cos \frac{L\Delta \beta _{C}}{2}-i\frac{1-\alpha ^{2}}{1+\alpha ^{2}}\sin 
\frac{L\Delta \beta _{C}}{2} & \frac{2\alpha }{1+\alpha ^{2}}\sin \frac{%
L\Delta \beta _{C}}{2} \\ 
-\frac{2\alpha }{1+\alpha ^{2}}\sin \frac{L\Delta \beta _{C}}{2} & \cos 
\frac{L\Delta \beta _{C}}{2}+i\frac{1-\alpha ^{2}}{1+\alpha ^{2}}\sin \frac{%
L\Delta \beta _{C}}{2}%
\end{array}%
\right)
\end{equation*}%
with $\alpha =\frac{2\gamma }{n_{x}^{2}-n_{y}^{2}+\sqrt{\left(
n_{x}^{2}-n_{y}^{2}\right) ^{2}4\gamma ^{2}}}$ and $\gamma $ being the
nondiagonal term in the relative dielectric constant tensor $\left( 
\begin{array}{ccc}
n_{x}^{2} & i\gamma & 0 \\ 
-i\gamma & n_{y}^{2} & 0 \\ 
0 & 0 & n_{z}^{2}%
\end{array}%
\right) $.

Linear and circular birefringence allow for the implementation of any $%
U\left( 2\right) $ transformation in the polarization-encoded qubits.
Preparation and measurement of the polarization-encoded qubits is obtained
by polarizing fibers and polarizing beam-splitters.

\textbf{(2) }$LP_{11}$\textbf{\ coding}

For a fiber with a parabolic index profile, the $LP_{11}$ modes may be
approximated by the first harmonic excitations along the $x$ and $y$
directions. Denoting by $a^{\dagger }$ and $b^{\dagger }$ the corresponding
creation operators, one has the following correspondence 
\begin{equation}
\left\vert 0\right\rangle \leftrightarrow LP_{01};\left( a^{\dagger
}\left\vert 0\right\rangle ,b^{\dagger }\left\vert 0\right\rangle \right)
\leftrightarrow LP_{11}  \label{2.15}
\end{equation}%
The $SU\left( 2\right) $ group, operating irreducibly in the 2-dimensional
space $\left( a^{\dagger }\left\vert 0\right\rangle ,b^{\dagger }\left\vert
0\right\rangle \right) $, is the following subgroup of the Weyl-symplectic
group in 2-dimensions 
\begin{equation}
\begin{array}{lll}
J_{+} & = & a^{\dagger }b \\ 
J_{-} & = & b^{\dagger }a \\ 
J_{3} & = & \frac{1}{2}\left( a^{\dagger }a-b^{\dagger }b\right)%
\end{array}
\label{2.16}
\end{equation}%
As explained in Ref.\cite{Manko1} and as follows from Eq.(\ref{2.5}) in
Sect. 2.1, changing the index profile along $x$ and $y$ as well as the
coefficient of the Laplacian one has access to all generators of the
two-dimensional Weyl-symplectic group and in particular to those of the $%
SU\left( 2\right) $ subgroup. Therefore, by engineering the index profile,
all unitary rotations may be implemented on the $LP_{11}-$encoded qubits.

Requiring a precise adjustment of the index profile, an unitary manipulation
of the $LP_{11}-$encoded qubits is more complex than the corresponding
operation on polarization-encoded qubits. Therefore this encoding might be
only recommended for control qubits.

In the quantumlike scheme one deals not with single photon events, but with
steady-state beams. Therefore conversion between the two encodings is
relatively easy using standard optical techniques.

\subsection{Two-qubit gates}

To obtain universal computation, in addition to one-qubit gates performing
arbitrary unitary transformations, one needs at least one entangling gate.
This is a gate that, together with one-qubit gates, generates all $U\left(
4\right) $ transformations. The CNOT, CS (controlled sign) or CP (controlled
phase) gates are such gates, but there are many others (Appendix A).

\subsubsection{A two-qubit controlled gate using LP$_{11}$ coding}

\begin{figure}[htb]
\centering
\includegraphics[width=0.6\textwidth]{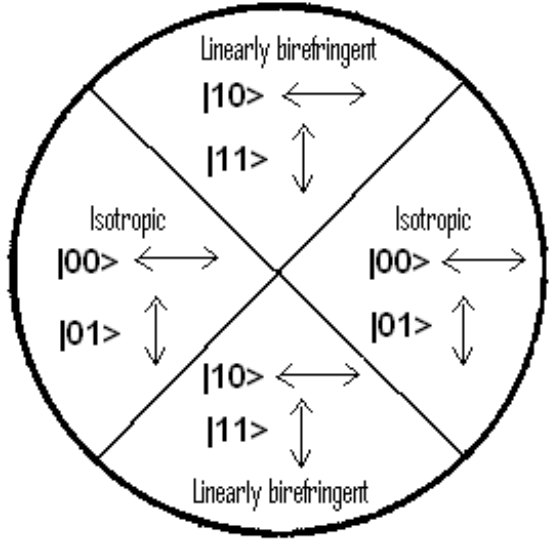}
\caption{Coding of a
controlled gate, using $LP_{11}$ modes for the control qubit and
polarization for the target qubit}
\label{Fig2}
\end{figure}

Here one shows how to obtain a controlled (entangling) gate using the two
qubit codings discussed before. On a fiber carrying $LP_{11}$ modes, the $%
LP_{11}$ mode has four degrees of freedom, two of them associated to the two
possible orientations of the mode (see Fig.\ref{Fig1}) and the other two to
the polarization. Let the two orientations of the $LP_{11}$ mode code the
control qubit and the polarization code the target qubit. For later
convenience the codes for the $\left\vert 1\right\rangle $ and$\ \left\vert
0\right\rangle $ qubits will be $V,H$ (vertical, horizontal) for the
polarizations (target) and $a,b$ for the positions (control) of the $LP_{11}$
modes. If the fiber is constructed in such a way that the $\left\vert
1\right\rangle $ sectors in the $LP_{11}$ mode are linearly birefringent and
the $\left\vert 0\right\rangle $ sectors are isotropic (see Fig.\ref{Fig2}),
a phase gate is obtained corresponding to the matrix 
\begin{equation}
M=\left( 
\begin{array}{cccc}
1 & 0 & 0 & 0 \\ 
0 & 1 & 0 & 0 \\ 
0 & 0 & 1 & 0 \\ 
0 & 0 & 0 & e^{i\theta }%
\end{array}%
\right)  \label{2.20}
\end{equation}%
in the basis $\left( \left\vert 00\right\rangle ,\left\vert 01\right\rangle
,\left\vert 10\right\rangle ,\left\vert 11\right\rangle \right) $, the first
entry being the control qubit and the second the target qubit. $\theta $ is
the additional phase that the $\left\vert 1\right\rangle _{t}$ target qubit
obtains in the $\left\vert 1\right\rangle _{c}$ sector of the control qubit.
In all cases there is a global phase that should be taken into account
arising from the $z-$propagation in the gate. With different choices of the
birefringence distribution other entangling $U\left( 4\right) $ matrices may
be obtained. 

Suppose that at the input of the gate the beam is a superposition of the $%
LP_{11}$ modes polarized on the $x,y$ plane ($\alpha _{1}\left\vert
0\right\rangle _{t}+\alpha _{2}\left\vert 1\right\rangle _{t}$) and that it
is the position (control) mode $a$ that is active. Then in the sector $a$ of
the fiber the output is%
\begin{equation}
\left\vert 1\right\rangle _{c}\otimes \left( \alpha _{1}\left\vert
0\right\rangle _{t}+\alpha _{2}e^{i\theta }\left\vert 1\right\rangle
_{t}\right) =\alpha _{1}\left\vert 10\right\rangle +\alpha _{2}e^{i\theta
}\left\vert 11\right\rangle ,  \label{2.21}
\end{equation}%
whereas in the $b$ sector the target qubit is unchanged. That is, the
degrees of freedom of the beam are entangled.

The nature of this entanglement\footnote{%
Some authors have claimed that the notion of entanglement should include
other features in addition to non-separability. Here entanglement is simply
used in the sense of non-separability.} is what has been called \textit{%
local entanglement} in the sense that it refers to the degrees of freedom
carried by the same physical entity. For a more general control qubit ($%
\beta _{1}\left\vert 0\right\rangle _{c}+\beta _{2}\left\vert 1\right\rangle
_{c}$) one has%
\begin{equation}
\beta _{1}\left\vert 0\right\rangle _{c}\otimes \left( \alpha _{1}\left\vert
0\right\rangle _{t}+\alpha _{2}\left\vert 1\right\rangle _{t}\right) +\beta
_{2}\left\vert 1\right\rangle _{c}\otimes \left( \alpha _{1}\left\vert
0\right\rangle _{t}+\alpha _{2}e^{i\theta }\left\vert 1\right\rangle
_{t}\right)  \label{2.22}
\end{equation}%
which would be faithfully implemented in the $LP_{11}$ gate. The target
qubit changes but only in the sector $a$ of the gate.

The usual statement that entangling two-qubit gates requires a nonlinear
effect, actually refers to the tensor product in (\ref{2.22}), which here is
obtained by local entanglement. The local entanglement, here associated to
the sharing of degrees of freedom by the same physical system, is, after
all, not so very different from the \textit{nonlocal entanglement} in
quantum mechanics. In quantum mechanics two photons may become entangled if
they have interacted in the past, in general because they were produced by a
common source\footnote{%
They may also be entangled by \textit{entanglement swapping} which involves
measurement, a nonlinear operation.}. They then share a common wavefunction
and, in this sense they are also parts of the same physical system. They
only become independent entities if the wavefunction decoheres, and then
entanglement is gone. So local and nonlocal entanglement are not so very
different as it might seem. On this optical entanglement of the beam degrees
of freedom there is another parallel with quantum mechanics. In quantum
mechanics the more noteworthy feature of entanglement is the fact that
correlation between the photons remains if at a later time they are well
separated in space. Here the role of time is played by the longitudinal $z-$%
coordinate of the fiber and the entanglement that occurs in the gate may be
observed at a later $z$. This, of course, if noise or the fiber
imperfections do not destroy space coherence. Like in quantum mechanics. In
short, entanglement requires interaction and remembrance of the interaction
effects along the propagation path.

In some quantum computing applications, for example in quantum Fourier
transform (QFT) as will be seen later, the full entangled output of the
phase gates is not used. Instead, in each line of the output of the QFT one
would want to find%
\begin{equation*}
\beta _{1}\left( \alpha _{1}\left\vert 0\right\rangle _{t}+\alpha
_{2}\left\vert 1\right\rangle _{t}\right) +\beta _{2}\left( \alpha
_{1}\left\vert 0\right\rangle _{t}+\alpha _{2}e^{i\theta }\left\vert
1\right\rangle _{t}\right) ,
\end{equation*}%
that is, a partial trace over the control qubit is effectively done.

If instead of linearly birefringency the $\left\vert 1\bullet \right\rangle $
region is circularly birefringent, also entangling gates may be constructed.
\ Here the two-bit gate is based on the four degrees of freedom of the $%
LP_{11}$ modes of a circular fiber. A similar construction might done using
the $TE,TM-12$ modes of a rectangular fiber. Modern fiber optics technology
is also able to handle multimode fibers which would provide entangling gates
for many more qubits.

Instead of a single fiber carrying $LP_{11}$ modes, one may use two fibers
(or light wave guides on a chip) one for the control position code $a$ and
the other for the code $b$. Each one of the light guides might carry the
full polarization information or the $a-$fiber might only contain the
vertical ($V$) component and the $b-$fiber the horizontal ($H$) component.
For future reference all these equivalent possibilities will be denoted as a 
$G-$gate.

Depending on its position on the quantum circuits, qubits may play the role
of target or control qubits. Therefore to each qubit one associates two
synchronous wave guides, to carry both position and polarization
information. While one of the lines carries optically the full polarization,
the other might well be electrical, with the interaction of polarization $%
\left( V,H\right) $ and position $\left( a,b\right) $ modes carried out by
optical or electro-optical means. Notice also that conversion of
polarization to position and vice versa is easily obtained by polarizing
beam splitters and polarization preserving fibers. The main challenge in
this dual coding scheme is to preserve linearity in the gate. In a
controlled phase gate only the $b-$line needs to enter the gate, the
polarization coming from the target line being established in this line
which is then passed through the appropriate retarder.

\subsubsection{Two-qubit gates with polarization coding}

\begin{figure}[htb]
\centering
\includegraphics[width=0.6\textwidth]{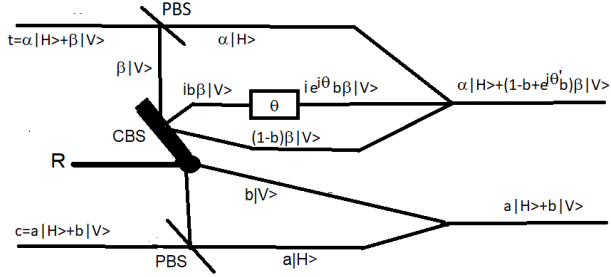}
\caption{Optical
two qubit phase gate with polarization coding. PBS = polarizing beam splitter; CBS = controlled beam splitter; R = reference beam. $\theta ^{\prime }=\protect\theta +\frac{\protect\pi}{2}$}
\label{OCBS}
\end{figure}

A different alternative for the construction of two-qubit gates would be to
use only one type of coding, for example polarization coding. In this case
the tensor product of control and target qubits is not achieved by the
coupling position-polarization, but it requires an interaction between the
two polarized beams, which only occurs through interaction with an optical
active medium. Fig.\ref{OCBS} sketches the required mechanism. After being
split by a polarizing beam splitter (PBS) the $V$ component of the target
beam is further split by another unit (controlled beam splitter, CBS) that
is controlled by the $V$ component of the control beam. One of the branches
is then passed through a phase retarder ($\theta $) to implement the
controlled phase $2-$qubit gate. This implements the operation in Eq.(\ref%
{2.23}). The essential element is the controlled beam splitter (CBS) which
can be achieved by a dynamical holography mechanism. A grating, dynamically
created on a material by interaction of the control and a reference beam,
splits the target beam. Optically and electro-optically controlled beam
splitters have been discussed and constructed before (see for example \cite%
{Sio} - \cite{ZografPRE}\ and references in \cite{Saleh},\cite{Li}). However
they operate mostly in an ON-OFF regime and here, as seen in Eq.\ref{2.23},
one needs linear operation. In Appendix B, the basic theory of one such
device is discussed as well as the requirements and challenges faced to
obtain linear operation.

\begin{equation}
\begin{array}{ccc}
\left. 
\begin{array}{c}
\alpha _{1}\left\vert H\right\rangle _{t}+\alpha _{2}\left\vert
V\right\rangle _{t} \\ 
\beta _{1}\left\vert H\right\rangle _{c}+\beta _{2}\left\vert V\right\rangle
_{c}%
\end{array}%
\right\} & \rightarrow & \beta _{1}\alpha _{1}\left\vert
H_{c}H_{t}\right\rangle \oplus \beta _{1}\alpha _{2}\left\vert
H_{c}V_{t}\right\rangle \oplus \beta _{2}\alpha _{1}\left\vert
V_{c}H_{t}\right\rangle \oplus \beta _{2}\alpha _{2}e^{i\theta }\left\vert
V_{c}V_{t}\right\rangle%
\end{array}
\label{2.23}
\end{equation}

\subsection{On the physical implementation of the gates}

As a general remark on the optical implementation of the operations of
quantum-like computing, it should be pointed out that one is in a more
favorable position than in the usual one-photon quantum computing
implementation. Here one deals with light beams and therefore nonlinear
effects are much easier to obtain. Furthermore one deals not with a
transient temporal phenomenon, but with the establishment, in a optical
network, of a steady state phenomenon. The initial state at the input of the
quantum-like circuit must be established by a coherent source which also
acts as a reference beam at other points of the circuit. The role of time
being played by a particular space coordinate, all the interference and gate
operations are performed until a steady state configuration is established
in the network, the final result of the calculation being read-off at some
well defined coordinate.

This also means that, as long as all superposition and interference
phenomena are implemented by optical waves, some intermediate gate
operations might be performed by electro-optical means. For example in a
controlled phase gate the amplitude and phase of the vertical polarizations
of control and target beams may be measured by heterodyning with the
reference beam and then, with the result of the gate operation computed by
electronic means, the same reference beam might by the appropriate retarders
generate the optical output beams. Also at intermediate points of the
network the signals may even be split, examined or amplified as long as the
phase is preserved or the phase change is duly taken into account. Of course
all-optical operation of the gates and of the whole circuit is desirable and
a goal to be achieved.

There is, in these intermediate measurements, no conflict with the no
cloning theorem of quantum information. In the usual proof of the no cloning
theorem, one assumes that an unitary operator $U$ exists such that $U|\psi
0>=|\psi \psi >$ for all $\psi $ and then, by applying $U$ to $\gamma
=\alpha \phi _{1}+\beta \phi _{2}$ obtain $U|\gamma 0>=\alpha |\phi _{1}\phi
_{1}>+\beta |\phi _{2}\phi _{2}>\neq |\gamma \gamma >$, a contradiction. No
cloning means that, given an unknown quantum state, no measurement can find
out what was exactly its wave function before the measurement. By contrast
given a beam of light one can split it in a polarization basis by a
polarizing beam splitter and then by heterodyning it with a coherent
reference beam find the amplitude and phase of each one of the components.
Given that knowledge, and because the phase is defined module $2\pi $, the
beam may then be synchronously reproduced.

In conclusion: the possibility to measure and then reproduce the
quantum-like signal, means that it will not be appropriate for cryptography
purposes. However, because it may have interference, parallelism, (local)
entanglement and unitary propagation along a (computing) coordinate, it may
be used for computation purposes.

\subsection{Nonlinear gates}

In the previous subsections the emphasis has been on linear gates, because
they are the ones most useful for computation purposes. However quantum
technology is not only quantum computing and nonlinear quantum (or
quantum-like) effects are also of interest. The electric field associated to
a single photon is very weak. This poses a major problem for all-optical
quantum operations using single photons, because significant,
medium-mediated, non-linear interactions would be required between two
photons. A very strong cooperative effect of atoms would be required to
perform interaction of single-photon signals. The Kerr effect at the one
photon level might be enhanced by choosing frequencies near resonances of
the material, but then appreciable loss effects would be expected.

In the optical quantum-like approach the signals, being coded not with
single photons but with light beams, nonlinear effects are much easier to
obtain. In particular, a great development has already been achieved with
nonlinear effects for switching purposes in classical all-optical networks.
Directional couplers are used as optical switches, as power dividers\ or
combiners, multiplexers, demultiplexers and intensity modulators. On-off
logic gates based on the Kerr effect have also been proposed by several
authors.

First studied by Jensen \cite{Jensen} the nonlinear directional coupler is a
robust device exploring the Kerr effect. In spite of its nonlinear nature,
by exploring the role of constants of motion, an analytic solution may be
obtained for the input-output transfer function of the device \cite{VilelaOC}%
. Therefore a precise quantitative control of the transfer function is
obtained. For the reader convenience, the main equations and parameters of
the coupler are summarized in the Appendix C. Denoting by $\overrightarrow{E}%
^{(1)}\left( 0\right) ,\overrightarrow{E}^{(2)}\left( 0\right) ,%
\overrightarrow{E}^{(1)}\left( L\right) ,\overrightarrow{E}^{(2)}\left(
L\right) $ the transversal electric fields at the input and output of the
two ports of a coupler ($1$ and $2$) of length $L$, one has a transfer
function%
\begin{equation}
\begin{array}{lll}
E_{j}^{(1)}\left( L\right)  & = & \frac{1}{2}\left\{ 
\begin{array}{c}
\left( e^{iL\overset{-}{\beta }^{(+)}}M^{(+)}+e^{iL\overset{-}{\beta }%
^{(-)}}M^{(-)}\right) _{jk}E_{k}^{(1)}\left( 0\right)  \\ 
+\left( e^{iL\overset{-}{\beta }^{(+)}}M^{(+)}-e^{iL\overset{-}{\beta }%
^{(-)}}M^{(-)}\right) _{jk}E_{k}^{(2)}\left( 0\right) 
\end{array}%
\right\}  \\ 
E_{j}^{(2)}\left( L\right)  & = & \frac{1}{2}\left\{ 
\begin{array}{c}
\left( e^{iL\overset{-}{\beta }^{(+)}}M^{(+)}-e^{iz\overset{-}{L\beta }%
^{(-)}}M^{(-)}\right) _{jk}E_{k}^{(1)}\left( 0\right)  \\ 
+\left( e^{iL\overset{-}{\beta }^{(+)}}M^{(+)}+e^{iL\overset{-}{\beta }%
^{(-)}}M^{(-)}\right) _{jk}E_{k}^{(2)}\left( 0\right) 
\end{array}%
\right\} 
\end{array}
\label{NL1}
\end{equation}%
where the matrices $M^{(+)}$, $M^{(-)}$ and the propagation factors $\beta
^{(+)}$, $\beta ^{(-)}$ associated to the symmetric and asymmetric modes are
completely specified by the material parameters of the coupler (Eqs. \ref%
{B.15}, \ref{B.16}). Through the constants of motion they have a nonlinear
dependence on the coupler medium and on the intensity of the beams. Of
course in the linear case $M^{(+)}$and $M^{(-)}$ are unit matrices.

For practical purposes one should notice that propagating through the
coupler each beam suffers changes of phase and polarization rotations due
both to itself and to the signal in the other beam, this latter action being
the one that is more relevant for the computational effect of the device.
Many different nonlinear operations may be obtained by the appropriate
choice of the parameters.

\section{Quantum modules}

\subsection{Quantumlike Fourier transform with 2-qubit optical gates}

\begin{figure}[htb]
\centering
\includegraphics[width=0.6\textwidth]{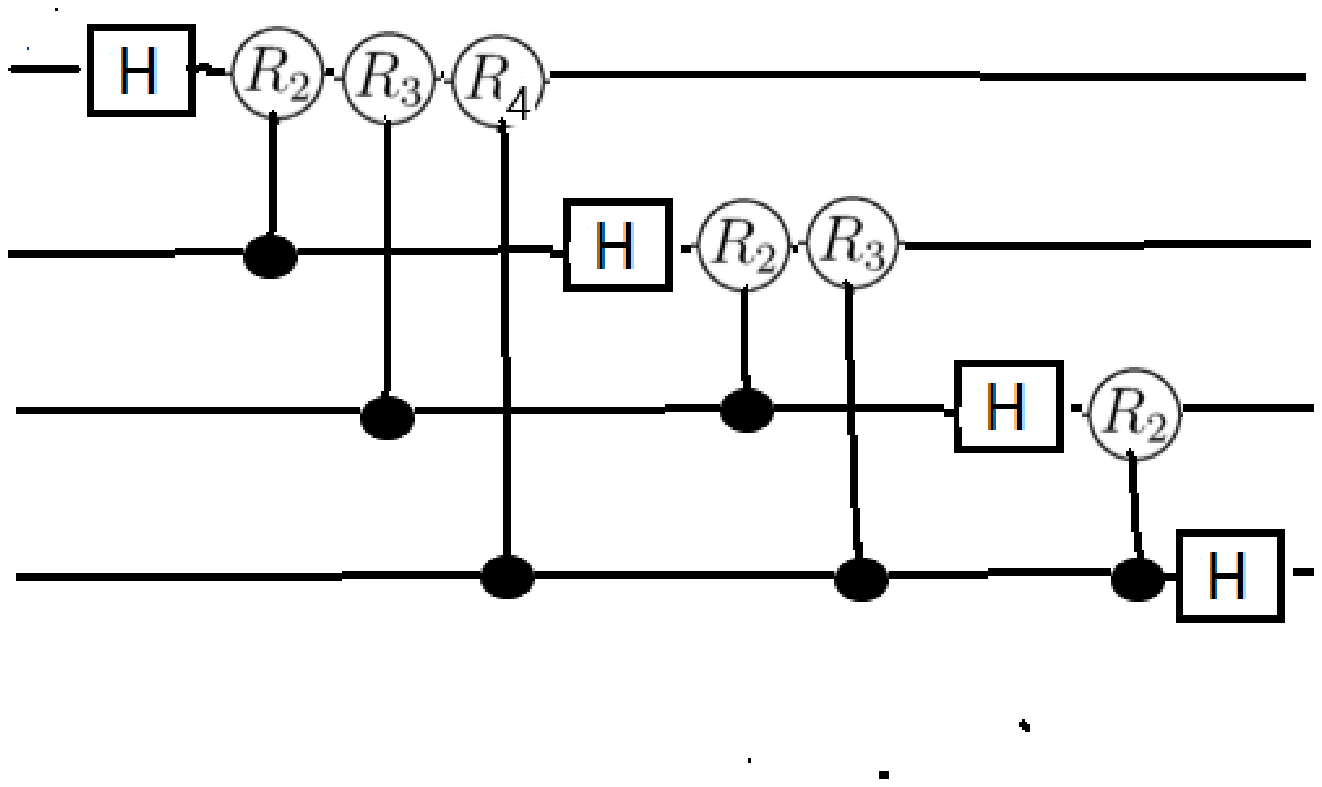}
\caption{A $O\left( 2n\right) $ quantum
Fourier transform circuit for 4 qubits}
\label{QFT_n2}
\end{figure}

A very important element in the quantum algorithms is the quantum Fourier
transform (QFT). For $n$ qubits and $N=2^{n}$ the QFT is%
\begin{equation}
y_{k}=\frac{1}{\sqrt{N}}\sum_{l=0}^{N-1}x_{l}e^{i2\pi lk/N},  \label{QFT1}
\end{equation}%
the $N$ number sets $\left\{ y\right\} $ and $\left\{ x\right\} $ being
coded by the $n$ qubits as follows%
\begin{equation}
x=\left( j_{1},j_{2},\cdots ,j_{n}\right) =j_{1}2^{n-1}+j_{2}2^{n-2}+\cdots
+j_{n}2^{0}  \label{QFT2}
\end{equation}%
The QFT may be looked at as an unitary transformation in the computational
basis of $n$ qubits, implementing the transformation \cite{Nielsen}%
\begin{eqnarray}
&&\left\lfloor j_{1}j_{2}\cdots j_{n}\right\rangle  \notag \\
&\rightarrow &\frac{1}{2^{n/2}}\left\{ \left( \left\lfloor 0\right\rangle
+e^{i2\pi \frac{j_{n}}{2}}\left\lfloor 1\right\rangle \right) \left(
\left\lfloor 0\right\rangle +e^{i2\pi \left( \frac{j_{n-1}}{2}+\frac{j_{n}}{4%
}\right) }\left\lfloor 1\right\rangle \right) \cdots \left( \left\lfloor
0\right\rangle +e^{i2\pi \left( \frac{j_{1}}{2}+\frac{j_{2}}{4}+\cdots +%
\frac{j_{n}}{2^{n}}\right) }\left\lfloor 1\right\rangle \right) \right\} 
\notag \\
&&  \label{QFT3}
\end{eqnarray}%
This decomposition of the QFT leads directly to the quantum circuit (for 4
qubits) in Fig.\ref{QFT_n2} where $H$ and $R_{k}$ are the Hadamard and the
controlled phase gates%
\begin{equation}
H=\left( 
\begin{array}{cc}
1 & 1 \\ 
1 & -1%
\end{array}%
\right) ;\;R_{k}=\left( 
\begin{array}{cc}
1 & 0 \\ 
0 & e^{i2\pi /2^{k}}%
\end{array}%
\right)  \label{QFT4}
\end{equation}%

This circuit has $n(n+1)/2$ gates which, exploring the
non-conflicting simultaneous application of the gates, may be implemented in 
$O\left( 2n\right) $ steps. There are however more efficient wirings \cite%
{fast1} - \cite{Moore2}.

Griffiths and Niu \cite{Niu} have proposed a semiclassical approach to the
quantum Fourier transform. It is semiclassical in the sense that it requires
a measurements of the output qubits to obtain a signal to control the gates.
In the time evolution approach to quantum computing this scheme would only
be applicable when the QFT is the final step in the quantum circuit. However
in the quantum-like approach because, as discussed before, measured beams
may be fully restored, the Griffiths and Niu configuration may be used at
any point in the circuit.

\begin{figure}[htb]
\centering
\includegraphics[width=0.6\textwidth]{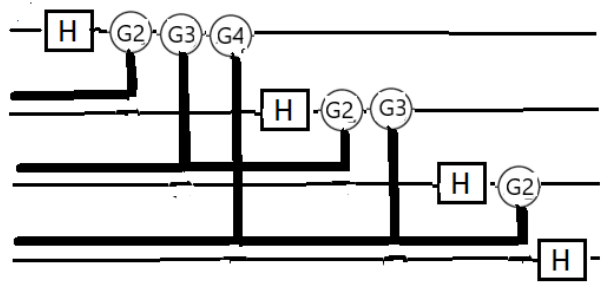}
\caption{Quantum-like Fourier transform using $LP_{11}
$ coding}
\label{QFT_On2}
\end{figure}

When using a single coding in the optical gates, for example polarization
coding, the QFT circuits for quantum-like computation would be identical to
the classical ones. However, when the $LP_{11}$ coding scheme (with one or
two wave guides) is used, the configuration might be slightly different. Fig.%
\ref{QFT_On2} displays one such implementation. In each input, except the
first, the input qubits are duplicated, assigned both to the polarization
modes of single mode ($LP_{01}$) fiber and to position $LP_{11}$ modes. The $%
H-$modules are Hadamard gates implemented by one-qubit gates with $LP_{01}$
modes polarization. Both the polarization ($V,H$) and the position ($a,b$)
information are fed to the gate. There the $a,b$ information and the
polarization ($V,H$) are used to generate a polarized $LP_{11}$ signal which
is fed to a partially birefringent fiber, which implements a two-qubit phase
gate, as described in Section 2.4.1. Notice that whereas the polarization
information is naturally carried in a $LP_{01}$ mode (the fine lines in Fig.%
\ref{QFT_On2}) the position information (the thick lines) for the $LP_{11}$
mode may be carried to the gate electronically or by a $LP_{11}$ fiber,
whatever is more convenient. At the end of the polarized $LP_{11}$ fiber in
the gate, the output polarization is obtained by merging the $a$ and $b$
modes into a $LP_{01}$ polarized mode. All gates are identical, differing
only on the length of the birefringent $LP_{11}$ fiber section. 

\subsection{Optical gateless quantum-like Fourier transforms}

The form of the quantum Fourier transform (Eq.\ref{QFT1}) is formally
identical to the classical discrete Fourier transform (DFT). In this sense,
what the QFT does is a DFT on the amplitudes of the quantum state. On the
other hand it is known that the Fourier transform may be obtained from a
light front, representing the function, by observation of the far field (or
the focused far field) at several angles. This led to several optical
proposals for the DFT by, for example, passing a coherent light through a
zero or $\pi $ phase mask and observing the far field in the focus plane of
a lens. These purely optical approaches that have also been proposed \cite%
{Tomita} - \cite{Macfaden} for the QFT with single photons may, even more
easily, be adapted to the light beam quantum-like approach.

\subsection{Oracles}

Oracles \cite{Vedral} \cite{Gilyen} are functions%
\begin{equation*}
f:\left( 0,1\right) ^{m}\rightarrow \left( 0,1\right) ^{n}
\end{equation*}%
which, typically, are needed both for the preparation of the input signal to
the quantum circuit and for queries about the final state. In terms of a
polarization coding of beams in the quantum-like approach, these are
functions%
\begin{equation*}
f:\left( H,V\right) ^{m}\rightarrow \left( H,V\right) ^{n}
\end{equation*}%
Such functions may be implemented by linear couplers, beam splitters,
interferometers, phase rotaters and the two-qubit gates discussed before. It
is desirable to use electro-optical control in these units to have
programmable flexibility of the oracles.

\section{Conclusions}

1) In this paper (and in \cite{Manko1}) by identifying a Schr\"{o}%
dinger-like evolution along a space coordinate of a classical system, we
have concluded that quantum computation might be carried out both by quantum
systems evolving in time and by a classical wave system evolving along a
space coordinate. This steals the primacy of quantum systems to execute
quantum computing operations. Even more, one might say that quantum
computing is more general than quantum mechanics or simply that in quantum
mechanics Nature is doing quantum computing along the time direction.

2) There is, of course, a difference in these two modalities of quantum
computing due to the particular nature of our observer status in the
universe for which, to look at a timeline (at a particular space) has
properties distinct from looking at a spaceline (at a particular time). When
looking at a timeline, after the operation the same time is no longer there,
in contrast with the timely permanence of a spaceline. As a result if a
measurement is made with a projection filter in the space evolution, the
same results are obtained as in quantum mechanics, but on the other hand
there are alternative ways to observe which give complete access to the
value of the wave function.

3) The optical implementations of the one and two qubit gates in this paper
have been kept are simple as possible, using only $LP_{01}$ and $LP_{11}$
modes. However with the growing sophistication on handling multimode fibers
it is conceivable that, using this optical quantum-like approach, it will be
possible to obtain high degrees of circuit compactness and parallelism. Of
particular interest for the development of interesting quantum-like devices
are the recent technological advances in space light modulators (SLM) \cite%
{Pinho} \cite{ParkJ}.

4) The current and potential applications of quantum technology are not
restricted to quantum computing, other promising uses are in fields of
control and communications. Whereas it seems that in quantum computing the
linear gates are the most useful, nonlinear gates are expected to be
potentially useful in other applications. This was the main motivation to
discuss at some length in section 2 and in the appendix C\ the analytical
aspects of the nonlinear circuits.

5) As stated before, there are, in addition to light waves, other systems
which display quantum-like behavior when its evolution along a space
coordinate is observed. Not all of them will be as appropriate as light to
perform computations, in particular because of the need to maintain
coherence in the evolution. Nevertheless a case that might deserve some
attention is the case of spin waves \cite{Csaba} \cite{Chumak}.

\section{Appendix A. Entangling two-qubit gates}

It is known \cite{Barenco} that arbitrary one-qubit gates together with a
two-qubit CNOT are capable of universal quantum computation. It the follows
that, more generally, any two-qubit gate capable of generating, together
with the one-qubit gates, the full $U(4)$ group would also be universal.
Such two-qubit gates have been called \textit{entangling} (or imprimitive)
gates, because they map decomposable states into indecomposable ones. A gate
that is not entangling is called \textit{primitive }\cite{Brylinski}.

Let $e_{ij}$ be a $4\times 4$ matrix with elements 
\begin{equation}
\left( e_{ij}\right) _{mn}=\delta _{im}\delta _{jn}  \label{A.1}
\end{equation}%
Then, the $16$ Lie algebra generators of $U(4)$ are 
\begin{equation}
\begin{array}{lll}
I_{ij} & = & i\left( e_{ij}-e_{ji}\right) \\ 
J_{ij} & = & e_{ij}+e_{ji}\qquad i\neq j \\ 
e_{ii} &  & 
\end{array}
\label{A.2}
\end{equation}%
They are related to the Lie algebra generators of $U\left( 2\right) \otimes
U\left( 2\right) $ by 
\begin{equation}
\sigma _{\mu }\otimes \sigma _{\nu }=\left( 
\begin{array}{cccc}
\sum_{i}e_{ii} & J_{12}+J_{34} & -I_{12}-I_{34} & e_{11}-e_{22}+e_{33}-e_{44}
\\ 
J_{13}+J_{24} & J_{14}+J_{23} & -I_{14}+I_{23} & J_{13}-J_{24} \\ 
-I_{13}-I_{24} & -I_{14}-I_{23} & J_{14}-J_{23} & -I_{13}+I_{24} \\ 
e_{11}+e_{22}-e_{33}-e_{44} & J_{12}-J_{34} & -I_{12}+I_{34} & 
e_{11}-e_{22}-e_{33}+e_{44}%
\end{array}%
\right)  \label{A.3}
\end{equation}%
where $\sigma _{\mu }=\left\{ \sigma _{0}\equiv \mathbf{1},\sigma
_{1},\sigma _{2},\sigma _{3}\right\} $ are the identity $2\times 2$ matrix
and the Pauli matrices.

The elements in the first line and the first column of the matrix in (\ref%
{A.3}), namely $\mathbf{1}\otimes \sigma _{\nu }$ and $\sigma _{\mu }\otimes 
\mathbf{1}$, are the algebraic elements associated to one-qubit operations.
The remaining $9$ elements in (\ref{A.3}) are of the form $\sigma
_{i}\otimes \sigma _{j}$ ($i,j=1,2,3$). From the commutators 
\begin{eqnarray}
\left[ \mathbf{1}\otimes \sigma _{i},\sigma _{a}\otimes \sigma _{b}\right]
&=&\sigma _{a}\otimes \left[ \sigma _{i},\sigma _{b}\right]  \label{A.4} \\
\left[ \sigma _{i}\otimes \mathbf{1,}\sigma _{a}\otimes \sigma _{b}\right]
&=&\left[ \sigma _{i},\sigma _{a}\right] \otimes \sigma _{b}  \notag
\end{eqnarray}%
it follows that, given any one of the $9$ elements $\sigma _{i}\otimes
\sigma _{j}$ it is possible to generate the full $U\left( 4\right) $ algebra
by commutation with the (one-qubit) generators $\mathbf{1}\otimes \sigma
_{\nu }$ and $\sigma _{\mu }\otimes \mathbf{1}$. These $9$ elements are
therefore a basis for the imprimitive (entangling) elements of the algebra.
Linear combinations of these elements as well as linear combinations with
one-qubit transformations are also entangling.

\section{Appendix B. An optically controlled beam splitter}

Many controllable beam splitters have been proposed in the past. They use
either mechanical displacement of metasurfaces \cite{WangOE},
electro-optical modulators and a Mach-Zehnder interferometer \cite{MaOSA},
optical bistability by surface plasmons \cite{SongOPTIK}, etc.

Optically controlled beam splitters have been discussed. For example \cite%
{Sio} uses a grating made of polymer slices alternated with layers of
aligned nematic liquid crystal. When the liquid crystal is aligned the input
light beam is split into a transmitted and a refracted component, however
when another pump beam is turned on, the liquid crystal suffers a nematic to
isotropic phase transition, the refractive index contrast vanishes and the
structure becomes transparent to the incoming light. Because fine-tuning of
the index contrast seems difficult, this interesting device is mostly suited
for an ON-OFF operation mode. The same applies to electro-optic operated
liquid crystal devices \cite{ZografPRE}.

The ON-OFF behavior of the controlled beam splitters is appropriate for
digital communication purposes, but for analog or quantum-like computing
applications a smoother, linear or quasi-linear, dependence on the control
signal is desirable. Quantum-like applications are even more demanding
because information on the phase of the control signal should be taken into
account.

The propagation of a transversal electric field in a nonlinear media is
described by the equation%
\begin{equation}
\triangle \mathbb{E}-\mu _{0}\varepsilon _{0}\frac{\partial ^{2}\mathbb{E}}{%
\partial t^{2}}=\mu _{0}\frac{\partial ^{2}P_{L}}{\partial t^{2}}+\mu _{0}%
\frac{\partial ^{2}P_{NL}}{\partial t^{2}}  \label{B1}
\end{equation}%
For the nonlinear contribution to the refraction index one considers either
a Kerr or a photorefractive medium. Let 
\begin{equation}
\mathbb{E}=\overrightarrow{E}\left( x,z\right) e^{i\omega t}  \label{B2}
\end{equation}%
$(x,z)$ being the coordinates of the propagation plane of field. Consider a
fixed thick sinusoidal grating along the $x$ coordinate (Fig.\ref%
{fixed_grating}), and the propagation of a light wave on this grating, that
is%
\begin{equation}
\triangle \overrightarrow{E}\left( x,z\right) +\mu _{0}\varepsilon
_{0}\omega ^{2}\overrightarrow{E}\left( x,z\right) =-\mu _{0}\varepsilon
_{0}\chi ^{(1)}\omega ^{2}\overrightarrow{E}\left( x,z\right) -\mu
_{0}\varepsilon _{0}\chi ^{(NL)}\cos \left( Qx\right) \omega ^{2}%
\overrightarrow{E}\left( x,z\right)  \label{B3}
\end{equation}%

\begin{figure}[htb]
\centering
\includegraphics[width=0.6\textwidth]{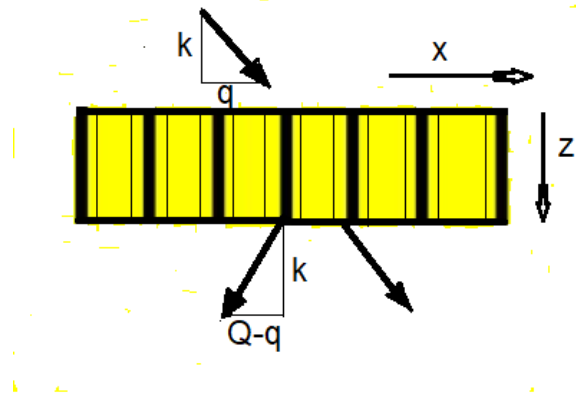}
\caption{A thick grating}
\label{fixed_grating}
\end{figure}

Passing to the Fourier
transform on the $x$ coordinate and writing%
\begin{equation}
\overrightarrow{E}\left( q,z\right) =\overrightarrow{\psi }\left( q,z\right)
e^{ikz}  \label{B4}
\end{equation}%
with $\overrightarrow{\psi }\left( q,z\right) $ having a slow variation on $%
z $, one obtains%
\begin{equation}
\left( -k^{2}-q^{2}\right) \overrightarrow{\psi }\left( q,z\right)
+2ik\partial _{z}\overrightarrow{\psi }\left( q,z\right) +\alpha 
\overrightarrow{\psi }\left( q,z\right) +\frac{\beta }{2}\left( 
\overrightarrow{\psi }\left( Q+q,z\right) +\overrightarrow{\psi }\left(
Q-q,z\right) \right) \simeq 0  \label{B5}
\end{equation}%
with%
\begin{eqnarray}
\alpha &=&\mu _{0}\varepsilon _{0}\chi ^{(1)}\omega ^{2}  \notag \\
\beta &=&\mu _{0}\varepsilon _{0}\chi ^{(NL)}\omega ^{2}  \label{B6}
\end{eqnarray}%
being the linear and nonlinear refractive coefficients. In Eq.(\ref{B5}) one
has neglected the term $\partial _{z}^{2}\overrightarrow{\psi }$.

With $q$ and $Q>0$, constructive interference of the diffractive components
in the thick grated slab requires%
\begin{eqnarray}
k^{2}+q^{2} &=&\alpha   \notag \\
k^{2}+\left( Q-q\right) ^{2} &=&\alpha   \label{B7}
\end{eqnarray}%
When only the $0th$ and the $1st$ diffraction orders are non-evanescent, $%
Q=2q$ and one has the following equations for the transmitted and diffracted
components%
\begin{eqnarray}
2ik\frac{\partial \overrightarrow{\psi }\left( q,z\right) }{\partial z} &=&-%
\frac{\beta }{2}\overrightarrow{\psi }\left( Q-q,z\right)   \notag \\
2ik\frac{\partial \overrightarrow{\psi }\left( Q-q,z\right) }{\partial z}
&=&-\frac{\beta }{2}\overrightarrow{\psi }\left( q,z\right)   \label{B8}
\end{eqnarray}%
with solution%
\begin{eqnarray}
\overrightarrow{\psi }\left( q,z\right)  &=&\overrightarrow{\psi }\left(
q,0\right) \cos \left( \frac{\beta }{4k}z\right)   \notag \\
\overrightarrow{\psi }\left( Q-q,z\right)  &=&i\overrightarrow{\psi }\left(
q,0\right) \sin \left( \frac{\beta }{4k}z\right)   \label{B9}
\end{eqnarray}%
In conclusion: the amount of splitting of the beam by the thick grating is
controlled by the nonlinear contribution to the refractive index.

Now, to have the splitting of the beam controlled by another light beam, the
grating should not be fixed but created by the intensity of the other beam.
Because one wants to have the tuning to be also a function of the phase,
consider $3$ light beams (target, control and reference, $t,c,R$)%
\begin{eqnarray}
\overrightarrow{E}_{t} &=&\left\vert \overrightarrow{E}_{t}\right\vert 
\overrightarrow{\epsilon }_{t}e^{i\left( \omega t-\overrightarrow{k_{t}}%
\cdot \overrightarrow{x}+\theta _{t}\right) }  \notag \\
\overrightarrow{E}_{c} &=&\left\vert \overrightarrow{E}_{c}\right\vert 
\overrightarrow{\epsilon }_{c}e^{i\left( \omega t-\overrightarrow{k_{c}}%
\cdot \overrightarrow{x}+\theta _{c}\right) }  \notag \\
\overrightarrow{E}_{R} &=&\left\vert \overrightarrow{E}_{R}\right\vert 
\overrightarrow{\epsilon }_{R}e^{i\left( \omega t-\overrightarrow{k_{R}}%
\cdot \overrightarrow{x}+\theta _{R}\right) }  \label{B10}
\end{eqnarray}%
The intensity of the sum of the three signals is%
\begin{eqnarray}
\left\vert \overrightarrow{E}_{t}\overrightarrow{+E}_{c}+\overrightarrow{E}%
_{R}\right\vert ^{2} &=&\left\vert \overrightarrow{E}_{t}\right\vert
^{2}+\left\vert \overrightarrow{E}_{c}\right\vert ^{2}+\left\vert 
\overrightarrow{E}_{R}\right\vert ^{2}  \notag \\
&&+\overrightarrow{\epsilon }_{t}\cdot \overrightarrow{\epsilon }%
_{c}\left\vert \overrightarrow{E}_{t}\right\vert \left\vert \overrightarrow{E%
}_{c}\right\vert \left( e^{i\left( \theta _{c}-\theta _{t}+\left( 
\overrightarrow{k_{t}}-\overrightarrow{k_{c}}\right) \cdot \overrightarrow{x}%
\right) }+c.c.\right)   \notag \\
&&+\overrightarrow{\epsilon }_{t}\cdot \overrightarrow{\epsilon }%
_{R}\left\vert \overrightarrow{E}_{t}\right\vert \left\vert \overrightarrow{E%
}_{R}\right\vert \left( e^{i\left( \theta _{t}-\theta _{R}+\left( 
\overrightarrow{k_{R}}-\overrightarrow{k_{t}}\right) \cdot \overrightarrow{x}%
\right) }+c.c.\right)   \notag \\
&&+\overrightarrow{\epsilon }_{c}\cdot \overrightarrow{\epsilon }%
_{R}\left\vert \overrightarrow{E}_{R}\right\vert \left\vert \overrightarrow{E%
}_{c}\right\vert \left( e^{i\left( \theta _{c}-\theta _{R}+\left( 
\overrightarrow{k_{R}}-\overrightarrow{k_{c}}\right) \cdot \overrightarrow{x}%
\right) }+c.c.\right)   \notag \\
&&  \label{B11}
\end{eqnarray}%
With linearly polarized signals it is always possible to have%
\begin{eqnarray}
\overrightarrow{\epsilon }_{t}\cdot \overrightarrow{\epsilon }_{c} &=&%
\overrightarrow{\epsilon }_{t}\cdot \overrightarrow{\epsilon }_{R}=0  \notag
\\
\overrightarrow{\epsilon }_{c}\cdot \overrightarrow{\epsilon }_{R} &\neq &0
\label{B12}
\end{eqnarray}%
Then, only the last of the mixed terms is nonvanishing and,%
\begin{equation}
\left\vert \overrightarrow{E}_{t}\overrightarrow{+E}_{c}+\overrightarrow{E}%
_{R}\right\vert ^{2}=\left\vert \overrightarrow{E}_{t}\right\vert
^{2}+\left\vert \overrightarrow{E}_{c}\right\vert ^{2}+\left\vert 
\overrightarrow{E}_{R}\right\vert ^{2}+\overrightarrow{\epsilon }_{c}\cdot 
\overrightarrow{\epsilon }_{R}\left\vert \overrightarrow{E}_{R}\right\vert
\left\vert \overrightarrow{E}_{c}\right\vert \left( e^{i\left( \theta
_{c}-\theta _{R}\right) }e^{i\left( \overrightarrow{k_{R}}-\overrightarrow{%
k_{c}}\right) \cdot \overrightarrow{x}}+c.c.\right)   \label{B13}
\end{equation}%
By the Kerr effect or on a photorefractive material, one may use this
intensity to create a grating along the $\overrightarrow{k_{R}}-%
\overrightarrow{k_{c}}$ direction. This holographic-like pattern carries the
information on the intensity and phase of the target signal, which with the
choice (\ref{B12}) is not contaminated by the interaction with the target
signal nor by the interaction of the reference beam with the target. The
intensity of the reference beam, in general larger than the one of the other
signals defines the amplitude of the grating effect.

For Kerr materials%
\begin{equation}
P_{NL}=\chi ^{(3)}\left\vert \overrightarrow{E}_{t}\overrightarrow{+E}_{c}+%
\overrightarrow{E}_{R}\right\vert ^{2}\left( \overrightarrow{E}_{t}%
\overrightarrow{+E}_{c}+\overrightarrow{E}_{R}\right)  \label{B14}
\end{equation}%
therefore the $\beta $ factor in Eq.(\ref{B9}) is proportional to $%
\overrightarrow{\epsilon }_{c}\cdot \overrightarrow{\epsilon }_{R}\left\vert 
\overrightarrow{E}_{c}\right\vert e^{i\theta _{c}}$, that is, the splitting
of the target beam would be directly controlled by the control beam. For
small $\beta $ this action is approximately linear on the amplitude of the
control, however deviations from linearity occur for large $\beta $. The
situation might be improved by manipulation of the $\overrightarrow{\epsilon 
}_{c}\cdot \overrightarrow{\epsilon }_{R}$ term, that is, making the control
beam pass through a medium that rotates the polarization as a function of
the intensity. Alternatively one might act on the intensity of the control
beam by electro-optical means to obtain $\beta =\sin ^{-1}\left( \alpha
\left\vert \overrightarrow{E}_{c}\right\vert \right) $ $.$

For photorefractive materials the change of the refractive index is
proportional to the space derivative of the intensity (see for example \cite%
{Saleh} ch. 21.4) and the mechanism is quite similar.

\section{Appendix C. Nonlinear directional couplers}

Directional couplers are useful devices currently used in fiber optics
communications. Because of the interaction between the two input fibers,
power fed into one fiber is transferred to the other. The amount of power
transfer is controlled by the coupling constant, the interaction length or
the phase mismatch between the inputs. If, in addition the material in the
coupler region has strong nonlinearity properties, the power transfer will
also depend on the intensities of the signals \cite{Jensen} \cite{Silberberg}%
. A large number of interesting effects take place in nonlinear directional
couplers \cite{Stegeman} \cite{Stegeman2} \cite{Kenis} \cite{Liu} with, in
particular, the possibility of performing all classical logic operations by
purely optical means \cite{Wang}.

Here one summarizes how, by exploring the constants of motion of the coupler
equation, explicit analytical solutions are obtained for both the linear and
nonlinear couplers, as used in Sect.2 for two-qubit gates. Further details
may be found in Ref. \cite{VilelaOC}.

Consider two linear optical fibers coming together into a coupler of
nonlinear material. The equation for the electric field is 
\begin{equation}
\triangle E-\mu _{0}\varepsilon _{0}\frac{\partial ^{2}E}{\partial t^{2}}%
=\mu _{0}\frac{\partial ^{2}P_{L}}{\partial t^{2}}+\mu _{0}\frac{\partial
^{2}P_{NL}}{\partial t^{2}},  \label{B.1}
\end{equation}%
$P_{L}\left( r,t\right) =\varepsilon _{0}\chi ^{(1)}E\left( r,t\right) $
being the linear polarization of the medium, $P_{NL}\left( r,t\right)
=\varepsilon _{0}\chi ^{(3)}\left\vert E\left( r,t\right) \right\vert
^{2}E\left( r,t\right) $ the nonlinear polarization in the instantaneous
nonlinear response approximation and transversal dependence of $\chi ^{(1)}$
and $\chi ^{(3)}$ have been considered negligible.

Separating fast and slow (time) variations 
\begin{equation}
\begin{array}{lll}
E\left( r,t\right) & = & \frac{1}{2}\left\{ \mathcal{E}\left( r,t\right)
e^{-i\omega _{0}t}+c.c.\right\} \\ 
P_{NL}\left( r,t\right) & = & \frac{1}{2}\left\{ \mathcal{P}_{NL}\left(
r,t\right) e^{-i\omega _{0}t}+c.c.\right\}%
\end{array}
\label{B.4}
\end{equation}
one obtains for the $e^{-i\omega _{0}t}$ part of a transversal mode 
\begin{equation}
P_{NL_{1,2}}\left( r,t\right) =\frac{3\varepsilon _{0}}{8}\chi ^{(3)}\left\{
e^{-i\omega _{0}t}\left[ \left( \left| \mathcal{E}_{1,2}\right| ^{2}+\frac{2%
}{3}\left| \mathcal{E}_{2,1}\right| ^{2}\right) \mathcal{E}_{1,2}+\frac{1}{3}%
\mathcal{E}_{2,1}\mathcal{E}_{2,1}\mathcal{E}_{1,2}^{*}\right] +c.c.\right\}
\label{B.5}
\end{equation}
the labels $1$ and $2$ denoting two orthogonal polarizations.

The dependence on transversal coordinates $\left( x,y\right) $ is separated
by considering 
\begin{equation}
E_{k}\left( r,t\right) =g\Psi _{k}^{(i)}\left( x,y,z\right) e^{i\beta
_{i}z}e^{-i\omega _{0}t}  \label{B.6}
\end{equation}%
with $\Psi _{k}^{(i)}\left( x,y,z\right) $ being an eigenmode of the coupler
with slow variation along $z$%
\begin{equation}
\Delta _{2}\Psi _{k}^{(i)}+\left( \frac{\omega _{0}^{2}}{c^{2}}\left( 1+\chi
^{(1)}\right) -\beta ^{(i)2}\right) \Psi _{k}^{(i)}=0  \label{B.7}
\end{equation}%
$\left( i\right) $ denotes the mode index, $k$ the polarization and $\Delta
_{2}=\left( \frac{\partial ^{2}}{\partial x^{2}}+\frac{\partial ^{2}}{%
\partial y^{2}}\right) $.

Neglecting $\frac{\partial ^{2}\Psi ^{(i)}}{\partial z^{2}}$ one obtains 
\begin{equation}
2i\beta ^{(i)}\frac{\partial \Psi _{1,2}^{(i)}}{\partial z}=-\frac{3\omega
_{0}^{2}}{4c^{2}}\chi ^{(3)}\left\{ \left( \left\vert \Psi
_{1,2}^{(i)}\right\vert ^{2}+\frac{2}{3}\left\vert \Psi
_{2,1}^{(i)}\right\vert ^{2}\right) \Psi _{1,2}^{(i)}+\frac{1}{3}\Psi
_{2,1}^{(i)}\Psi _{2,1}^{(i)}\Psi _{1,2}^{(i)\ast }\right\}  \label{B.8}
\end{equation}%
In the directional couplers the propagating beams are made to overlap along
one of the transversal coordinates $\left( x\right) $. Typically, in the
overlap region of the directional coupler, the eigenmodes are symmetric $%
\left( +\right) $ and antisymmetric $\left( -\right) $ functions on $x$, the
amplitudes in each fiber at the input and output of the coupler being
recovered by 
\begin{equation}
\begin{array}{lll}
\Psi _{k}^{(1)} & = & \frac{1}{2}\left( \Psi _{k}^{(+)}+\Psi
_{k}^{(-)}\right) \\ 
\Psi _{k}^{(2)} & = & \frac{1}{2}\left( \Psi _{k}^{(+)}-\Psi
_{k}^{(-)}\right)%
\end{array}
\label{B.9}
\end{equation}

An explicit analytic solution, also for the nonlinear coupler equation (\ref%
{B.8}), may be obtained by noticing that it has two constants of motion 
\begin{equation}
\begin{array}{lll}
\frac{\partial }{\partial z}\left\{ \left\vert \Psi _{1}^{(i)}\right\vert
^{2}+\left\vert \Psi _{2}^{(i)}\right\vert ^{2}\right\} & = & 0 \\ 
\frac{\partial }{\partial z}\left\{ \Psi _{1}^{(i)\ast }\Psi _{2}^{(i)}-\Psi
_{1}^{(i)}\Psi _{2}^{(i)\ast }\right\} & = & 0%
\end{array}
\label{B.10}
\end{equation}%
Therefore, defining 
\begin{equation}
\begin{array}{lll}
\left\vert \Psi _{1}^{(i)}\right\vert ^{2}+\left\vert \Psi
_{2}^{(i)}\right\vert ^{2} & = & \alpha ^{(i)} \\ 
\Psi _{1}^{(i)\ast }\Psi _{2}^{(i)}-\Psi _{1}^{(i)}\Psi _{2}^{(i)\ast } & =
& i\gamma ^{(i)}%
\end{array}
\label{B.11}
\end{equation}%
one obtains for the electrical field of the eigenmodes 
\begin{equation}
\begin{array}{lll}
i\frac{\partial E_{1}^{(i)}}{\partial z} & = & -\overset{-}{\beta }%
^{(i)}E_{1}^{(i)}-i\overset{-}{k}^{(i)}E_{2}^{(i)} \\ 
i\frac{\partial E_{2}^{(i)}}{\partial z} & = & -\overset{-}{\beta }%
^{(i)}E_{2}^{(i)}+i\overset{-}{k}^{(i)}E_{1}^{(i)}%
\end{array}
\label{B.12}
\end{equation}%
with 
\begin{equation}
\begin{array}{lll}
\overset{-}{\beta }^{(i)} & = & \beta ^{(i)}+\frac{3\omega _{0}^{2}}{8c^{2}}%
\frac{\chi ^{(3)}}{\beta ^{(i)}}\alpha ^{(i)} \\ 
\overset{-}{k}^{(i)} & = & \frac{\omega _{0}^{2}}{8c^{2}}\frac{\chi ^{(3)}}{%
\beta ^{(i)}}\gamma ^{(i)}%
\end{array}
\label{B.13}
\end{equation}%
Notice that, through $\alpha ^{(i)}$ and $\gamma ^{(i)}$, $\overset{-}{\beta 
}^{(i)}$ and $\overset{-}{k}^{(i)}$depend on the material properties, on the
geometry of the mode and also on its intensity. One may now obtain, for each
eigenmode, the input-output relation of the nonlinear coupler 
\begin{equation}
\begin{array}{lll}
E_{1}^{(i)}\left( z\right) & = & e^{i\overset{-}{\beta }^{(i)}z}\left\{
E_{1}^{(i)}\left( 0\right) \cos \left( \overset{-}{k}^{(i)}z\right)
-E_{2}^{(i)}\left( 0\right) \sin \left( \overset{-}{k}^{(i)}z\right) \right\}
\\ 
E_{2}^{(i)}\left( z\right) & = & e^{i\overset{-}{\beta }^{(i)}z}\left\{
E_{1}^{(i)}\left( 0\right) \sin \left( \overset{-}{k}^{(i)}z\right)
+E_{2}^{(i)}\left( 0\right) \cos \left( \overset{-}{k}^{(i)}z\right) \right\}%
\end{array}
\label{B.14}
\end{equation}%
the nonlinearity being embedded into $\overset{-}{\beta }^{(i)}$ and $%
\overset{-}{k}^{(i)}$%
\begin{equation}
\begin{array}{lll}
\overset{-}{\beta }^{(i)} & = & \beta ^{(i)}+\frac{3\omega _{0}^{2}}{8c^{2}}%
\frac{\chi ^{(3)}}{\beta ^{(i)}}\left( \left\vert E_{1}^{(i)}\left( 0\right)
\right\vert ^{2}+\left\vert E_{2}^{(i)}\left( 0\right) \right\vert
^{2}\right) \\ 
\overset{-}{k}^{(i)} & = & \frac{\omega _{0}^{2}}{4c^{2}}\frac{\chi ^{(3)}}{%
\beta ^{(i)}}\textnormal{Im}\left( E_{1}^{(i)\ast }\left( 0\right)
E_{2}^{(i)}\left( 0\right) \right)%
\end{array}
\label{B.15}
\end{equation}%
To obtain the corresponding input-output relations in the two fibers one
defines a matrix 
\begin{equation}
M^{(\pm )}\left( z\right) =\left( 
\begin{array}{ll}
\cos \left( \overset{-}{k}^{(\pm )}z\right) & -\sin \left( \overset{-}{k}%
^{(\pm )}z\right) \\ 
\sin \left( \overset{-}{k}^{(\pm )}z\right) & \cos \left( \overset{-}{k}%
^{(\pm )}z\right)%
\end{array}%
\right)  \label{B.16}
\end{equation}%
Eq.(\ref{B.14}) is rewritten 
\begin{equation}
E^{(\pm )}\left( z\right) =e^{i\overset{-}{z\beta }^{(\pm )}}M^{(\pm
)}\left( z\right) E^{(\pm )}\left( 0\right)  \label{B.17}
\end{equation}%
$z$ being the interaction length of the directional coupler. Using (\ref{B.9}%
) the fields at the output of the coupler are related to the input fields by%
\begin{equation}
\begin{array}{lll}
E_{j}^{(1)}\left( z\right) & = & \frac{1}{2}\left\{ 
\begin{array}{c}
\left( e^{iz\overset{-}{\beta }^{(+)}}M^{(+)}+e^{iz\overset{-}{\beta }%
^{(-)}}M^{(-)}\right) _{jk}E_{k}^{(1)}\left( 0\right) \\ 
+\left( e^{iz\overset{-}{\beta }^{(+)}}M^{(+)}-e^{iz\overset{-}{\beta }%
^{(-)}}M^{(-)}\right) _{jk}E_{k}^{(2)}\left( 0\right)%
\end{array}%
\right\} \\ 
E_{j}^{(2)}\left( z\right) & = & \frac{1}{2}\left\{ 
\begin{array}{c}
\left( e^{iz\overset{-}{\beta }^{(+)}}M^{(+)}-e^{iz\overset{-}{\beta }%
^{(-)}}M^{(-)}\right) _{jk}E_{k}^{(1)}\left( 0\right) \\ 
+\left( e^{iz\overset{-}{\beta }^{(+)}}M^{(+)}+e^{iz\overset{-}{\beta }%
^{(-)}}M^{(-)}\right) _{jk}E_{k}^{(2)}\left( 0\right)%
\end{array}%
\right\}%
\end{array}
\label{B.18}
\end{equation}%
For the linear coupler case the $M^{(\pm )}\left( z\right) $ matrices are
the unit matrices and the coupling arises only from the difference in the
propagation constants $\overset{-}{\beta }^{(+)},\overset{-}{\beta }^{(-)}$%
of symmetric and antisymmetric modes. However in both cases, linear and
nonlinear, explicit analytical expressions are obtained for the coupling as
a function of the input intensities and the material properties. In $\overset%
{-}{\beta }^{(i)}$the nonlinear effect is a function of the energy of the
incoming signals and $\overset{-}{k}^{(i)}$ has a geometrical interpretation
as%
\begin{equation*}
\overset{-}{k}^{(i)}=\frac{\omega _{0}^{2}}{8c^{2}}\frac{\chi ^{(3)}}{\beta
^{(i)}}\left\vert \Psi ^{(i)\ast }\times \Psi ^{(i)}\right\vert
\end{equation*}

Here it was assumed that the frequency of the two incoming signals to the
coupler is the same. If they have different frequencies $\omega _{1}$ and $%
\omega _{2}$ the corresponding constants of motion, as a function of the
associated fields $\Psi ^{1},\Psi ^{2}$, would be%
\begin{equation*}
\left\vert \Psi ^{1}\right\vert ^{2};\left\vert \Psi ^{2}\right\vert ^{2};%
\frac{\beta _{1}}{\omega _{1}^{2}}\Psi ^{1\ast }\times \Psi ^{1}+\frac{\beta
_{2}}{\omega _{2}^{2}}\Psi ^{2\ast }\times \Psi ^{2}
\end{equation*}%
However, in this case, these constants of motion do not seem to be
sufficient to obtain an explicit analytical solution.

\end{document}